\documentclass[aps, prb,amsmath,amssymb,twocolumn]{revtex4-1}
\usepackage[pdftex]{graphicx}
\usepackage{bm,braket,color}
\usepackage{hyperref}

\begin{document}
\title{Magnetoelectric Effects and Charge-Imbalanced Solenoids: \\Antiferro Quadrupole Orders in a Diamond Structure}

\author{Takayuki Ishitobi}
\author{Kazumasa Hattori}
\affiliation{Department of Physics, Tokyo Metropolitan University, Hachioji, Tokyo 192-0397, Japan}

%\date{\today}

\begin{abstract}
We study magnetoelectric (magneto-current) effects in a diamond structure under antiferro quadrupole (AFQ) orders. The AFQ orders break the spatial inversion symmetry and cause the current-induced magnetization. The current-induced magnetization strongly depends on the types of the order parameters and the direction of the current. This gives a way to the experimental identification of AFQ order parameters. We also discuss the current-induced magnetization under the AFQ orders in the diamond structure can be intuitively understood in terms of charge-imbalanced solenoids.
\end{abstract}

\maketitle

Spontaneous symmetry breaking and their impact on various responses in functional materials have attracted great interest in condensed matter physics\cite{Fiebig2005,Nagaosa2013,Edelstein1990}. Among various symmetries, a spatial inversion (SI) symmetry is one of the most basic symmetry as well as a time-reversal (TR) symmetry. Lack of a SI symmetry causes many interesting phenomena in variety of systems such as multiferroic materials\cite{Tokura2014}, noncentrosymmetric superconductors\cite{BauerSigrist2012}, and chiral crystals.\cite{FurukawaItou2017} In multiferroics, magnetic orders breaking the SI symmetry directly couple with electric degrees of freedom, i.e., atomic polarizations.\cite{Fiebig2005, Tokura2014} Controlling such cross correlations is expected to be useful for the future application to various high-performance devices\cite{Fiebig2016, Talebi2018}. 

Systems preserving the global SI symmetry but without local SI symmetry have been also attracted considerable attention.\cite{Fischer2011,Yanase2014, Hayami2014, Kudo2018}
They are called ``locally noncentrosymmetric systems''. The local symmetry around the active degrees of freedom lacks the local SI symmetry.
In such systems, novel classification of superconducting gap functions emerges\cite{Fischer2011} and a simple antiferro order can break the global SI symmetry.\cite{Yanase2014,Hayami2014} In this line of context, physics behind odd-parity and/or toroidal multipole moments is systematically discussed.\cite{WatanabeYanase2018,HayamiYatsushiroYanagiKusunose2018}

In a general framework of magnetoelectric (ME) effects\cite{Fiebig2005, Fiebig2016, Tokura2014},  linear ME effects in insulators need the SI and the TR symmetry breakings\cite{LandauLifshitzContinuousMedia}. In metals, however, the broken TR symmetry is not necessary, since the electric current induced by the  electric field can couple to, e.g., magnetizations, if the SI symmetry is broken. This is so-called the Edelstein effect\cite{Edelstein1990}, in other words, kinetic-magnetic effects, or magneto-current effects\cite{HayamiYatsushiroYanagiKusunose2018, WatanabeYanase2018, YodaYokoyamaMurakami2015}. 
The ME effects reflect not only the SI and TR symmetry, but also the other space group symmetry of the system\cite{HayamiYatsushiroYanagiKusunose2018, WatanabeYanase2018, HayamiKusunose2018}. 
Thus, the detailed analysis of the ME effects can be potential indicators for the symmetry of the system. 

In this Letter, we show that analyses of ME effects in 
the cage compounds Pr$T_{2}X_{20}$ ($T$=Ir, Rh, Ti, V, $X$=Zn, Al etc.)\cite{OnimaruKusunose2016} can be 
powerful tools for identifying their quadrupole order parameters. The Pr 1-2-20 systems
exhibit the quadrupole orders, two-channel Kondo effects, and superconductivity at low temperatures\cite{SakaiNakatsuji2011, Onimaru2011, SakaiKugaNakatsuji2012, Onimaru2012PRB}. 
In these systems, Pr$^{3+}$ ions with f$^2$ configurations form the diamond structure. The crystalline electronic field ground state under the $T_d$ point group is non-Kramers $\Gamma_3$ doublet with the electric quadrupole ($\mathcal{O}_{20}$, $\mathcal{O}_{22}$) and magnetic octupole (${\mathcal T}_{xyz}$) moments without magnetic dipole moments. 
Several theoretical studies have proposed various symmetry broken phases.\cite{HattoriTsunetsugu2014, HattoriTsunetsugu2016, Freyer2018,Lee2018, Patri2019}. 
Despite the intensive experimental and theoretical studies, their order parameters have not been fully understood. Concerning indirect evidences such as ultrasonic experiments\cite{Ishii2011, Koseki2011}, antiferro quadrupole (AFQ) orders are likely in PrIr$_2$Zn$_{20}$,\cite{Onimaru2011,Onimaru2012JPCM} PrRh$_2$Zn$_{20}$\cite{Onimaru2012PRB, Onimaru2012JPCM} and  PrV$_2$Al$_{20}$\cite{SakaiNakatsuji2011,Araki2013, Shimura2013, Ito2015}, while a ferro quadrupole (FQ) order in PrTi$_2$Al$_{20}$\cite{SakaiNakatsuji2011,Matsubayashi2012,SakaiKugaNakatsuji2012, Koseki2011, Sato2012, Taniguchi2016}. 

Under AFQ orders, the SI symmetry is broken, since the Pr ions do not locate at the inversion center in the diamond structure. Thus, current-induced magnetizations (CIM) can emerge. This must reflect the nature of the AFQ order parameters. Namely, $\mathcal{O}_{20}$ and $\mathcal{O}_{22}$ orders exhibit distinct responses in the ME effects and anisotropy in the current directions.
In this Letter, we will show that the CIM in a diamond structure under AFQ orders shows order parameter dependent anisotropy profile. We will also discuss the present ME effects can be intuitively understood by considering arrays of charge-imbalanced solenoids. 

 We consider $p$-orbital electrons interacting with quadrupole moments in a diamond structure. This particular choice of the orbital does not alter our main results, and thus, we will use this simple model throughout this Letter. 
Of course, the model is over simplified for discussing the Pr-based materials but our results are based on the symmetry argument, and thus, do not depend on the detail. 
 The $p$-orbital multiplets are split into $j=1/2$ and $j=3/2$ in the presence of spin-orbit interaction. We here retain the $j=3/2$ multiplet and this corresponds to $\Gamma_8$ state in the $T_d$ symmetry, since the
 $\Gamma_8$ orbital can directly couple with $\Gamma_3$ quadrupole moment ($\mathcal{O}_{20}$, $\mathcal{O}_{22}$)\cite{Cox1987}. 
  
 The quadrupole orders are represented by the one-body potential $V$, which originates from, e.g., quadrupolar Kondo coupling. 
 The one-body Hamiltonian matrix for the wavenumber $\bm k$  is given as 
 $H({\bm k})=H_0({\bm k})+V({\bm k})-\mu \tau_0\sigma_0\gamma_0$ with $\mu$ being the chemical potential and 
\begin{align}
	H_{0}({\bm k}) &= \sum_{\nu=0,x,y}  
(\mathcal{E}^{\nu}_{\bm{k}}\tau_{0}\sigma_{0} 
+ \bm{\eta}^{\nu}_{\bm{k}} \cdot \bm{\sigma} \tau_{y} 
+ \bm{d}^{\nu}_{\bm{k}}\cdot \bm{\tau}\sigma_{0} )\gamma_{\nu}, \label{eq:H0}\\
	V({\bm k})&=\sum_{\mu=x,z}\left(\Delta_{\rm AF}^\mu\tau_{\mu}\gamma_z+
	\Delta_{\rm F}^\mu\tau_{\mu}\right).\label{eq:V}
\end{align}
Here, 
$\bm{\sigma}$, $\bm{\tau}$, and $\bm{\gamma}$ are the Pauli matrices for 
the pseudospin, the orbital, and the sub-lattice degrees of freedom, respectively. 
$\sigma_0$, $\tau_0$, and $\gamma_0$ represent the identity matrix for the corresponding sector. The parameters 
$\mathcal{E}^{\nu}_{\bm{k}}$, $\bm{\eta}^{\nu}_{\bm{k}}$, and $\bm{d}^{\nu}_{\bm{k}}$ depend on the Slater Koster parameters and they are fixed as typical values; $(t^{nn}_{pp\sigma}, t^{nn}_{pp\pi}, t^{nnn}_{pp\sigma}, t^{nnn}_{pp\pi})=(1.0, -0.3, 0.5, -0.15)$, where the superscript $nn$ and $nnn$ denote the hopping between the nearest and 
next-nearest neighbors, respectively. In the following, $t^{nn}_{pp\sigma}$ for the $\sigma$ bonds is set to the energy unit. $\Delta_{\rm AF}^{z}(\Delta_{\rm AF}^{x})$ represents the uniform ${\mathcal O}_{20}^{\rm AF}$ (${\mathcal O}^{\rm AF}_{22}$) AFQ orders and 
$\Delta_{\rm F}^{z,x}$ represents the ferroic component ${\mathcal O}_{20,22}^{\rm F}$ generally induced in the 
AFQ states via the coupling $
\sim \Delta_{\rm F}^z[(\Delta_{\rm AF}^z)^2-(\Delta_{\rm AF}^x)^2]-
2\Delta_{\rm F}^x\Delta_{\rm AF}^z\Delta_{\rm AF}^x
$ allowed in $T_d$ symmetry\cite{HattoriTsunetsugu2014, HattoriTsunetsugu2016}. For simplicity, we will specify each phase by the primal order parameter ${\mathcal O}_{22,20}^{\rm AF}$. Diagonalizing $H(\bm{k})$ leads to eight eigenvalues $\epsilon_{n\bm{k}}$ and the Bloch functions (eigenvectors) $u_{n\bm{k}}$ for a given $\bm{k}$, where $n=1,2, \cdots, 8$ is the band index.

\begin{figure}[t!]
\begin{center}
\includegraphics[width=0.5\textwidth]{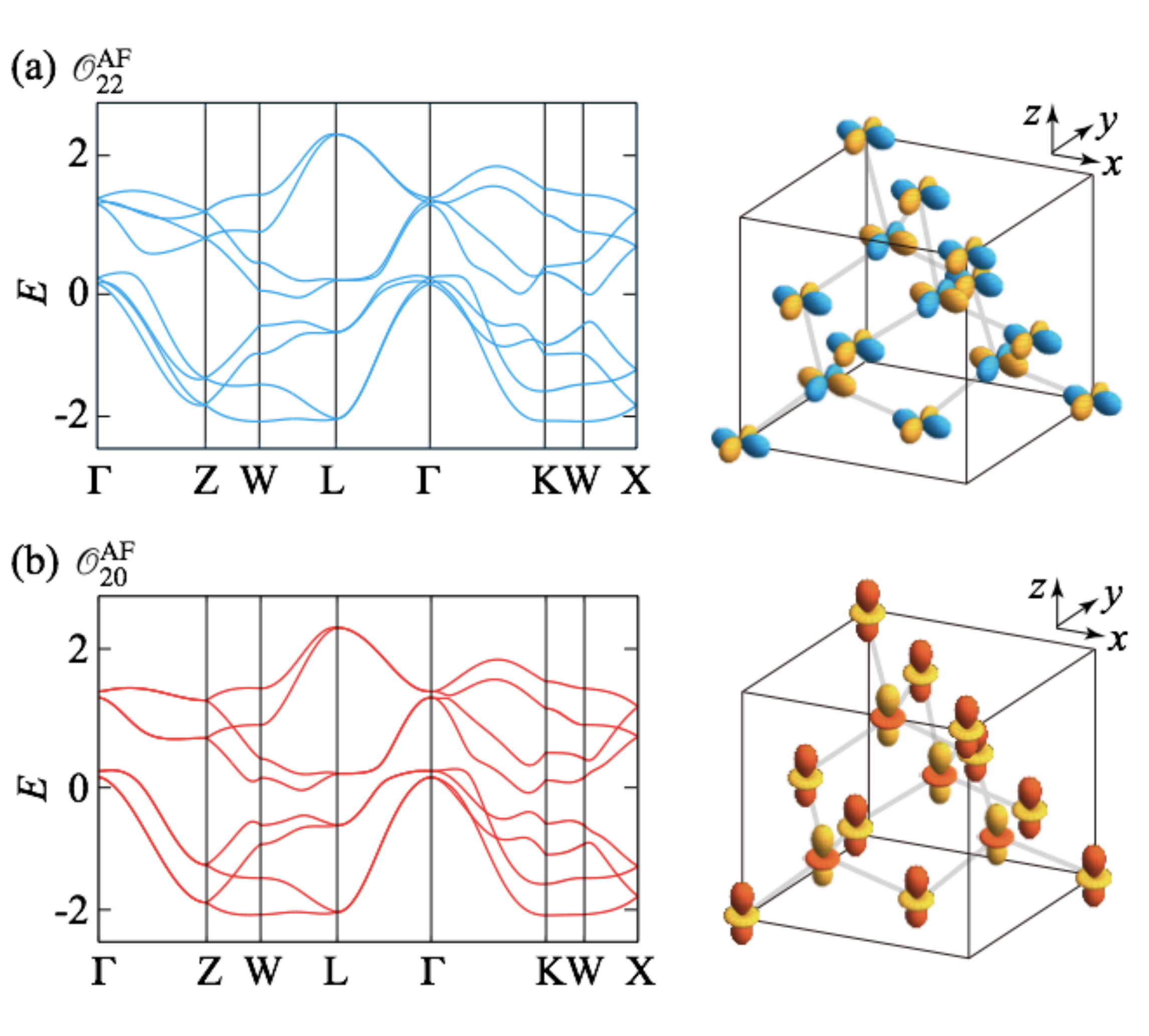}
\end{center}
\caption{\label{fig:band}(color online). The band structure for (a) ${\mathcal O}_{22}^{\rm AF}$ 
with $(\Delta_{\rm AF}^{z},\Delta_{\rm AF}^{x},\Delta_{\rm F}^{z},\Delta_{\rm F}^{x})=(0,0.3,0.05,0)$ and (b) ${\mathcal O}_{20}^{\rm AF}$ with $(0.3,0,0.05,0)$. A schematic view of the AFQ orders in a cubic unit cell are also shown.}
\end{figure}

Figures \ref{fig:band}(a) and (b) show the band structure under (a) $\mathcal{O}_{22}^{\rm AF}$ and for (b) $\mathcal{O}_{20}^{\rm AF}$ orders. 
Without $V$, every band is doubly degenerate with the space group $Fd\bar{3}m$ (No. 227), owing to the Kramers' theorem. When $V\ne 0$, the band structure is modified as determined by the symmetry of the quadrupole types, i.e. ${\mathcal O}_{20}$ or ${\mathcal O}_{22}$.
Along the $\Gamma$-$Z$ and $\Gamma$-$L$ lines, the degeneracy is lifted for (a) with the symmetry $I4_{1}22$ (No. 98), while is preserved for (b) with $P\bar{4}2m$ (No. 119). 
This is owing to the presence/absence of (110) mirror symmetry, which affects the anisotropy of the CIM as will be discussed later.

%%%%%%%%%%%%%%%%%%%%%%%%%%%%%%%%%%%%%%%%%%%%%%%%%%%%%%%%%%

%%%%%%%%%%%%%%%%%%%%%%%%%%%%%%%%%%%%%%%%%%%%%%%%%%%%%%%%%%

First, let us clarify the symmetry constraint 
for realizing CIM. In the diamond structure, the AFQ orders break the SI symmetry. The electric-current $\bm{\mathcal{J}}$, the magnetization $\bm{\mathcal M}$, and the AFQs couple as\cite{HayamiYatsushiroYanagiKusunose2018, WatanabeYanase2018}
\begin{align}
	&{\mathcal O}^{\rm AF}_{22}(2{\mathcal J}_z{\mathcal M}_z-{\mathcal J}_x{\mathcal M}_x-{\mathcal J}_y{\mathcal M}_y)\label{eq:lambda1},\\
	-&\sqrt{3}{\mathcal O}^{\rm AF}_{20}({\mathcal J}_x{\mathcal M}_x-{\mathcal J}_y{\mathcal M}_y).\label{eq:lambda2}
\end{align}
Comparing Eqs. (\ref{eq:lambda1}) and (\ref{eq:lambda2}), one can realize that the AFQs affect the anisotropy of the induced $\bm{\mathcal M}$ under finite $\bm{\mathcal J}$. For $\mathcal{O}^{\rm AF}_{22}$ order, the $z$ component of the magnetization ${\mathcal M}_z$ is induced by its parallel component ${\mathcal J}_z$, while no ${\mathcal M}_z$ is induced under $\mathcal{O}^{\rm AF}_{20}$ order. Note that Eq. (\ref{eq:lambda1}) is invariant when $x\leftrightarrow y$, while Eq. (\ref{eq:lambda2}) changes the sign for $x\leftrightarrow y$. Thus, for $\mathcal{O}^{\rm AF}_{20}$ order, the magnetization parallel to the current changes its sign when the current is rotated in the $xy$ plane. In contrast, it has no change for $\mathcal{O}^{\rm AF}_{22}$ order. 
By substituting magnetic field $\bm{H}$ for $\bm{\mathcal M}$, the AFQs also couple to $\bm{\mathcal J}$ and $\bm{H}$. This means that AFQ moments emerge when both $\bm{H}$ and $\bm{\mathcal J}$ are present, and their anisotropy reflects the couplings [Eqs (\ref{eq:lambda1}) and (\ref{eq:lambda2})]. This might cause an enhancement (suppression) of the transition temperature under the magnetic fields and the current. These properties can be used to identify the AFQ orders. 

We emphasize that the ME effect is one of direct tools for the identification of the AFQ order parameters. Note that the AFQ moments in the present system are classified as electric toroidal quadrupole (ETQ) moments\cite{HayamiYatsushiroYanagiKusunose2018} with $E_{u}^+$ representation in the $O_h$ symmetry (this is the symmetry at $\Gamma$ point). Here, ``+'' indicates the parity for the TR operation. Remember that natural conjugate fields for the toroidal moments are spatial derivative fields. For example, the magnetic toroidal dipoles couple to $\bm{\nabla}\times \bm{H}\sim\bm{\mathcal{J}}$\cite{Hayami2014, Zimmermann2014, SaitoAmitsuka2018}, while ETQ moments couple to $\sum_{\mu \nu} \lambda_{\mu \nu}X_{\mu}(\bm{\nabla}\times \bm{X})_{\nu}$, where $\bm{X}=\bm{H}$: the magneto-curernt effect\cite{Edelstein1990} or $\bm{X}=\bm{E}$ (electric field): the gyrotropic magnetic effect\cite{ZhongMooreSouza2016}, and $\lambda_{\mu \nu}$ is a symmetric traceless tensor.  Thus, the minimal ``field'' that couples to ETQ is $\sim \bm{H}(\bm{\nabla}\times \bm{H})\sim \bm{H}\bm{\mathcal{J}}$, and this is the simplest field for detecting the ETQ, i.e., the AFQ order parameters in this system. Recently, Hayami et al., have proposed that a bond-order ETQ state is realized in Cd$_2$Re$_2$O$_7$.\cite{Hayami2019}

Now, we will discuss two current-induced ``magnetizations''. One is the conventional magnetization $\bm{M}^{J}=g_{J}\mu_{B}\bm{J}$ originating from the atomic $j=3/2$ angular momentum $\bm{J}$ of the conduction electrons, where $\mu_{B}$ and $g_{J}$ is the Bohr magneton and the Lande's $g$-factor, respectively. The other is the orbital magnetization $\bm{M}^{{\rm orb}}$ from inter-site electron motions. The former is calculated via the linear response theory, while the latter is estimated by semi-classical Boltzmann formula\cite{ChangNiu1996, XiaoChangNiu2010, YodaYokoyamaMurakami2015} as
$
	M^{J/{\rm orb}}_{i}= \alpha^{J/{\rm orb}}_{ij} E_{j}
$, where $E_j$ is the $j$th component of the applied electric field $\bm{E}$ and 
\footnotesize
\begin{align}
\label{eq:alpha_J}
\alpha^{J}_{ij} =\frac{1}{N}\sum_{nm\bm{k}}  M^{Ji}_{nm,\bm{k}}\mathcal{J}_{mn,\bm{k}}^{j} \left[\frac{-i(f_{n\bm{k}}-f_{m\bm{k}})}{\omega_{nm,\bm{k}}^2+\delta^2} 
- \frac{f^{\prime}_{n\bm{k}}}{\delta}\Biggr|_{\omega_{nm,\bm{k}}=0} \right],
%- \frac{f^{\prime}_{n\bm{k}}}{\delta}\Biggr|_{\epsilon_{n\bm{k}}=\epsilon_{m\bm{k}}} \right],
%\alpha^{J}_{ij} = -i&\sum_{\bm{k}} \sum_{\epsilon_{n\bm{k}}\neq \epsilon_{m\bm{k}}} {M^{Ji}_{nm,\bm{k}}}\mathcal{J}_{mn,\bm{k}}^{j} \frac{(f_{n\bm{k}}-f_{m\bm{k}})}{\omega_{nm,\bm{k}}^2+\delta^2} \nonumber \\
%- &\sum_{\bm{k}}\sum_{\epsilon_{n\bm{k}}=\epsilon_{m\bm{k}}} {M^{Ji}_{nm,\bm{k}}}\mathcal{J}_{mn,\bm{k}}^{j} \frac{f^{\prime}_{n\bm{k}}}{\delta},
\end{align}
\begin{align}
\label{eq:alpha_orb}
\alpha^{{\rm orb}}_{ij} &= \frac{e}{N\delta}\sum_{n\bm{k}}  f^{\prime}_{n\bm{k}} \bigl\{ m^{i}_{n\bm{k}} + \frac{e}{\hbar}(\mu-\epsilon_{n\bm{k}}) \Omega^{i}_{n\bm{k}} \bigr\} v^{j}_{n\bm{k}}.
\end{align}
\normalsize
Here,  $N$ is the number of unit cells, $\omega_{nm,\bm{k}}=\epsilon_{n\bm{k}}-\epsilon_{m\bm{k}}$ and 
$\mathcal{J}^j_{mn,\bm{k}}$ represents the matrix elements of the  electric current $\bm{\mathcal{J}}=-e\partial H(\bm{k})/\partial \bm{k}$ for the $j$th component in the band basis and the notation is applied to the magnetization: $M^{Ji}_{nm,\bm{k}}$.  $f_{n\bm{k}}=f(\epsilon_{n\bm{k}})$ and $f^{\prime}_{n\bm{k}}$ are the Fermi distribution of the $n$th band and its  derivative with respect to the energy, respectively. 
$\delta$ is the temperature ($T$) independent impurity scattering rate. In Eq. (\ref{eq:alpha_orb}), $\bm{v}_{n\bm{k}}=\partial \epsilon_{n\bm{k}}/\partial \bm{k}$, $\bm{\Omega}_{n\bm{k}}=i\bm{\nabla} \times \braket{u_{n\bm{k}}|\bm{\nabla}|u_{n\bm{k}}}$ (the Berry curvature), and 
$\bm{m}_{n\bm{k}}=e{\rm Im} \braket{\bm{\nabla}u_{n\bm{k}}|\times(H(\bm{k})-\epsilon_{n\bm{k}})|\bm{\nabla}u_{n\bm{k}}}/(2\hbar)$: the orbital magnetic moment for  the $n$th band\cite{ChangNiu1996,  XiaoChangNiu2010}, where $e$ and $\hbar$ are the elementary charge and the Plank's constant, respectively. 
The first term in Eq. (\ref{eq:alpha_J}) is the electric-field-induced contribution, which vanishes in TR symmetric systems\cite{WatanabeYanase2017, HayamiYatsushiroYanagiKusunose2018}. The second term represents the current-induced contribution. In the present TR system, only the latter is finite. The first term in Eq. (\ref{eq:alpha_orb}) is owing to the local magnetic moments of the wave-packet\cite{ChangNiu1996, XiaoChangNiu2010}. The second term arises from the Berry curvature correction to the density of states\cite{ChangNiu1996, XiaoChangNiu2010}, although this vanishes at $T=0$\cite{YodaYokoyamaMurakami2015}.

%%%%%%%%%%%%%%%%%%%%%%%%%%%%%%%%%%%%%%%%%%%%%%%%%%%%%%%%%%

\begin{figure}[t!]
\vspace{-4mm}
\begin{center}
\includegraphics[angle=0,width=0.45\textwidth]{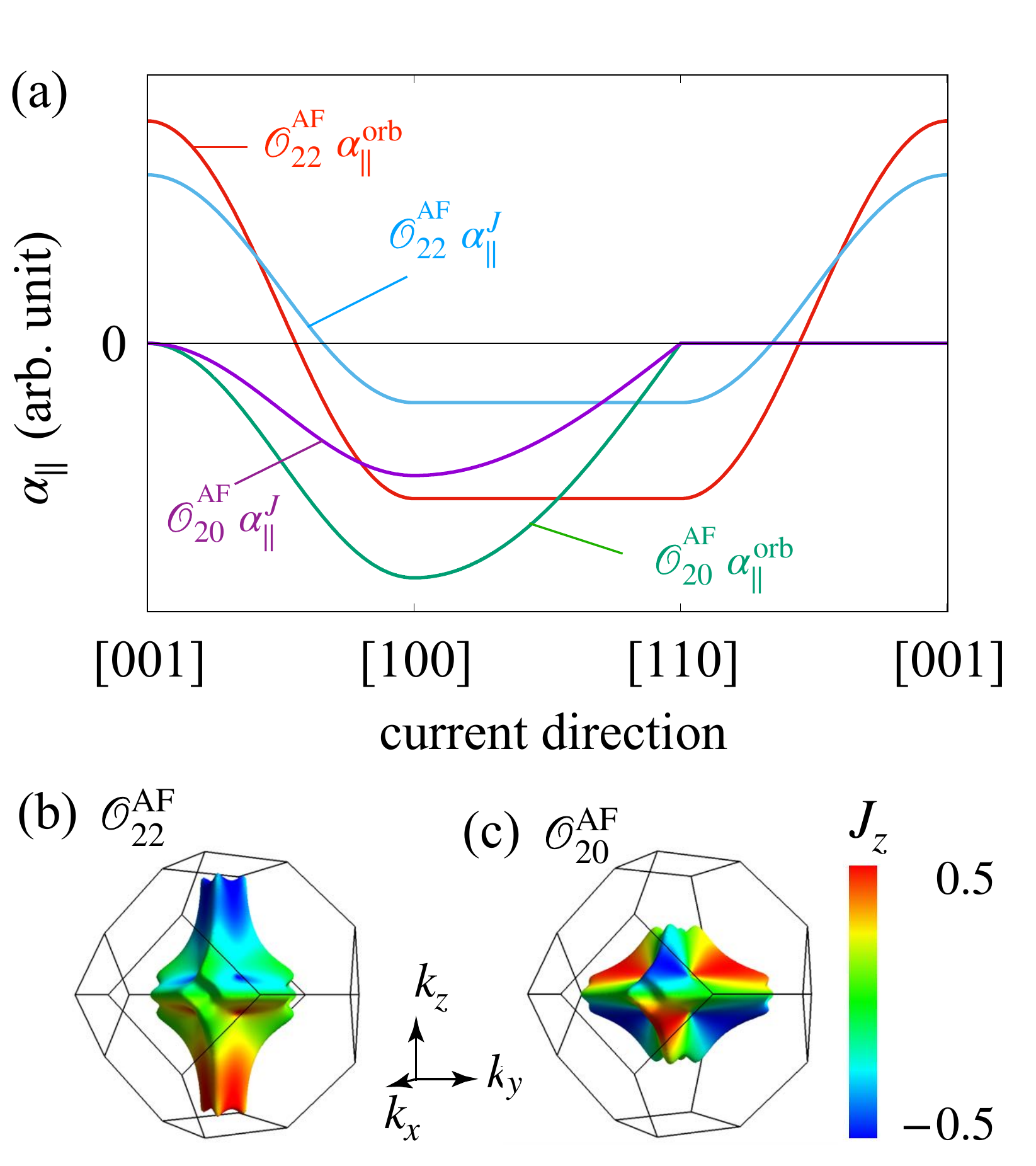}
\end{center}
\vspace{-4mm}
\caption{(color online). (a) The current direction dependence of $\alpha^{{\rm orb},J}_{\parallel}=\bm{E}\cdot \bm{M}^{{\rm orb},J}/|\bm{E}|^2$ for ${\mathcal O}_{22}^{\rm AF}$ and ${\mathcal O}_{20}^{\rm AF}$ orders with $\mu=1.0$. Main Fermi surface is drawn with the total angular momentum $J_{z}$ color-plotted for (b) $\mathcal{O}^{\rm AF}_{22}$ and (c) $\mathcal{O}^{\rm AF}_{20}$ orders. }
\label{fig:angle-dep}
\end{figure}

%%%%%%%%%%%%%%%%%%%%%%%%%%%%%%%%%%%%%%%%%%%%%%%%%%%%%%%%%%

Let us discuss the anisotropy of the CIM. It is useful to focus on the case of $\bm{\mathcal{M}} \parallel \bm{\mathcal{J}}$ in Eqs. (\ref{eq:alpha_J}) and (\ref{eq:alpha_orb}), and this component is denoted as $\alpha^{\rm orb}_{\parallel}$.  
Figure \ref{fig:angle-dep}(a) shows $\alpha^{\rm orb}_{\parallel}$ as a function of the current (electric field) direction. As expected from the symmetry arguments, for $\bm{\bm{E}}\parallel \hat{z}$, $|\bm{M}^{{\rm orb}}|=0$ for ${\mathcal O}_{20}^{\rm AF}$ order, while $|\bm{M}^{{\rm orb}}|>0$ for ${\mathcal O}_{22}^{\rm AF}$. Under ${\mathcal O}_{22}^{\rm AF}$ order, $\alpha^{\rm orb}_{\parallel}$ is constant on the $z=0$ plane,  which reflects the  symmetric coupling with respect to $x$ and $y$ [Eq. (\ref{eq:lambda1})]. For ${\mathcal O}_{20}^{\rm AF}$ order, such constant (even vanishing) behavior appears on the $x=y$ plane, owing to the antisymmetric coupling with respect to $x$ and $y$ [Eq. (\ref{eq:lambda2})]. 
Thus, one can realize that the magnetization measurement with rotating the current direction becomes a powerful tool for identifying the AFQ orders.
\if0
\begin{align}
 \lambda_{3} \{(\mathcal{O}_{22}^{AF})^3 -3(\mathcal{O}_{20}^{AF})^2 \mathcal{O}_{22}^{AF} \} \bm{\mathcal{J}} \cdot \bm{\mathcal{M}}. \label{eq:lambda3}
\end{align}
\fi
Note that the direction where the sign of $\alpha_\parallel^{\rm orb}$ changes on the $zx$ plane ($[001]$-$[100]$) for $\mathcal{O}_{22}^{\rm AF}$ is not the same as that for $\alpha^{J}_\parallel$, despite that they are expected to be the same, i.e., $[\sqrt{2}01]$ from Eqs. (\ref{eq:lambda1}) and (\ref{eq:lambda2}). This is owing to, e.g., the third-order coupling present in the ordered states.

The above anisotropic response can be understood by the magnetic moment distribution in $\bm{k}$ space for non-degenerate bands under the AFQ orders. To illustrate this, consider the case for $\bm{E} \parallel [001]$. Figures \ref{fig:angle-dep}(b) and \ref{fig:angle-dep}(c) show the distributions of the angular momentum $J_{z}$ on a main Fermi surface (FS) for $\mu=1.0$. 
For $\mathcal{O}^{\rm AF}_{22}$ order in Fig. \ref{fig:angle-dep}(b), the distribution reflects the linear coupling $k_{z} \mathcal{M}_{z}$. 
When $\bm{E} \parallel [001]$ is applied, the FS is shifted to $\hat{k}_{z}$ direction and a finite $\mathcal{M}_{z}$ appears. In contrast, the coupling is the Dresselhaus type $k_z(k_x^2-k_y^2)\mathcal{M}_{z}$ for $\mathcal{O}_{20}^{\rm AF}$ order 
in Fig. \ref{fig:angle-dep}(c). Thus, $\mathcal{M}_{z}$ is not induced by $\mathcal{J}_{z}$ for $\mathcal{O}_{20}^{\rm AF}$ order after integrated in the whole $\bm{k}$ space.

%%%%%%%%%%%%%%%%%%%%%%%%%%%%%%%%%%%%%%%%%%%%%%%%%%%%%%%%%%

\begin{figure}[t!]
\begin{center}
 \includegraphics[width=0.5\textwidth]{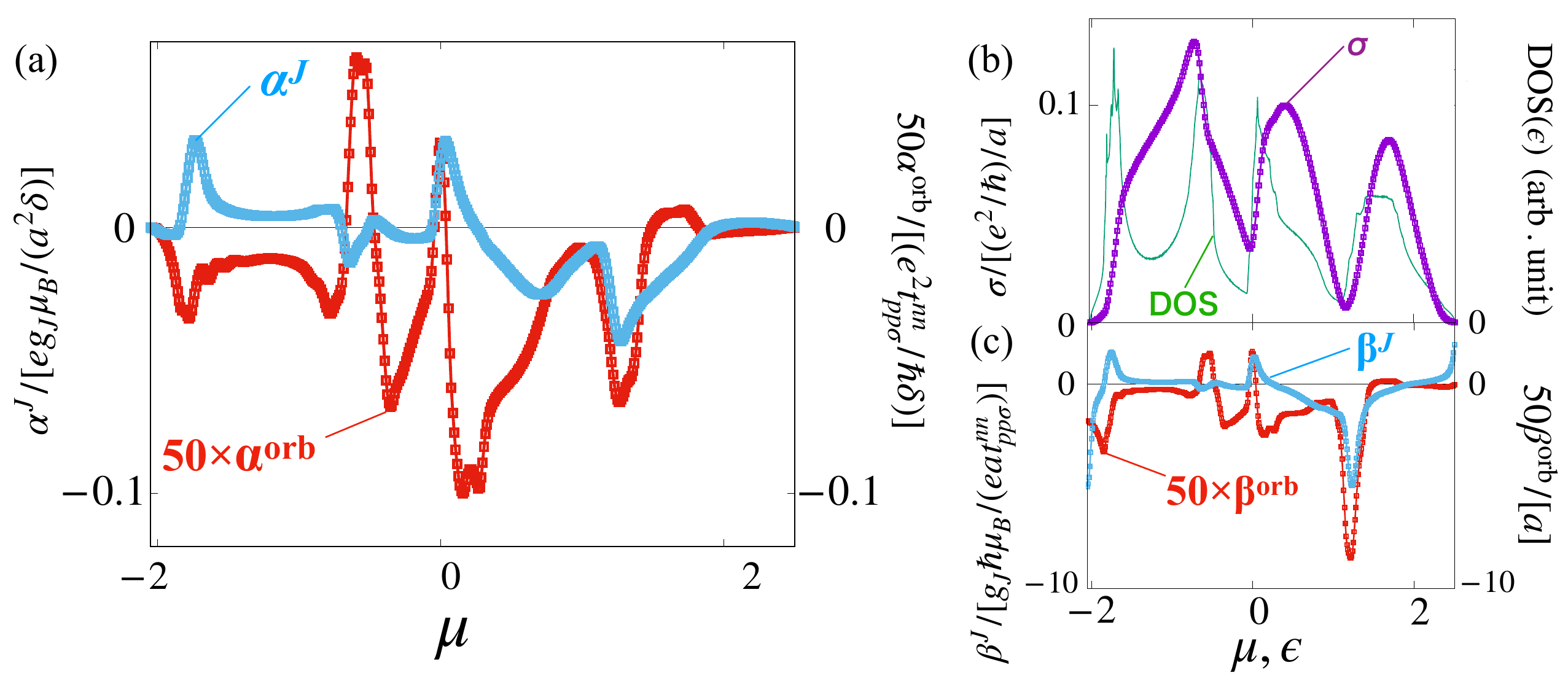}
\end{center}
\vspace{-4mm}
\caption{(color online). The chemical potential $\mu$ dependence of (a) the current-induced $\alpha^{J,{\rm orb}}_{xx}$, (b) the electric conductivity $\sigma_{xx}$ and DOS, and (c) $\beta^{J,{\rm orb}}_{xx}=M^{J,{\rm orb}}_{x}/j_{x}$ for $\mathcal{O}_{20}^{\rm AF}$ order under $\bm{E} \parallel [100]$ with $\Delta_{\rm AF}^{z}=0.012$, $\Delta_{\rm AF}^x=\Delta_{\rm F}^{z,x}=0$, and $T=0.001$.}
\label{fig:order}
\end{figure}

%%%%%%%%%%%%%%%%%%%%%%%%%%%%%%%%%%%%%%%%%%%%%%%%%%%%%%%%%%
 Now let us estimate the magnitude of CIM, considering Pr 1-2-20 systems in mind. First, consider the case for $t_{pp\sigma}^{nn}\sim 1$ eV, the lattice constant $a \sim 15$ \AA, $\Delta^{z,x}_{\rm AF}\sim 100$ K\cite{OnimaruKusunose2016}, and $\delta\sim 20$ K. Then, we find the conductivity $\sigma\sim 0.1/(\mu\Omega{\rm cm})$ consistent with a value of typical metal. For the ME effect, $\alpha^{J,{\rm orb}}_{xx}\sim 10^{-6}$ G/(V/m) and the induced moment per unit current density $\beta^{J,{\rm orb}}_{xx}\equiv M^{J,{\rm orb}}_x/j_x$ is $10^{-14}\mu_B$/(A/m$^2$) per site. The latter is much smaller than $\sim 10^{-10}\mu_B$/(A/m$^2$) observed in UNi$_4$B\cite{SaitoAmitsuka2018}. 
 
 	 Taking into account the renormalization owing to correlation effects, we set $\Delta_{\rm AF}^z/t_{pp\sigma}^{nn}=0.01$ and the results are shown in Figs. \ref{fig:order}(a)-(c). 
	 Here, we discuss ${\mathcal O_{20}}^{\rm AF}$ order, but similar results are obtained for ${\mathcal O_{22}}^{\rm AF}$ order except for the anisotropy. 
Variations of $\sigma$ is mostly owing to the density of states (DOS) as shown in Fig. \ref{fig:order}(b), while peaks of $\alpha_{xx}$ and $\beta_{xx}$ depend on the details of band structure, e.g., the band crossing points and the DOS. 
When we set $t_{pp\sigma}^{nn}=120$ K (thus, $\Delta^z_{\rm AF}=1$ K), $\delta=0.23$ K leads to $\sigma\sim 0.1/(\mu\Omega{\rm cm})$ [See Fig. \ref{fig:order}(b)]. 
	 For these parameters,  $\alpha^J_{xx}\sim 100\alpha^{\rm orb}_{xx}\sim 10^{-4}$ G/(V/m), and $\beta^J_{xx}\sim 10^{-12}\mu_B$/(A/m$^2$). For $j_x=10$ kA/m$^2$, we find $M_x\sim 10^{-8} \mu_B$ per site. 
	 
	 In Pr 1-2-20 systems, $f$-electrons are expected to contribute to the CIM, since they hybridize with the conduction electrons and acquire itinerancy. Note that while the ground state CEF state is nonmagnetic $\Gamma_3$, the itinerant part with the heavy mass has the magnetic dipole. We expect that correlation effects beyond the present analysis with $f$-electrons enhance CIM, as implied from the comparison between the theory\cite{Hayami2014} and experiments\cite{SaitoAmitsuka2018} in UNi$_4$B.
	 In Ref. 
\onlinecite{FurukawaItou2017}, 
$M\sim 10^{-8}\sim 10^{-9}\mu_B$ per site in Te is estimated by NMR. This is owing to the huge hyperfine coupling constant in $^{125}$Te. Thus, NMR is one of promising tools for the detection of CIM when the hyperfine coupling constant is large.
	 
	 We note that the larger $\Delta_{\rm AF}/$[band width], the larger induced magnetization emerges. Thus, materials with higher transition temperature and/or heavy mass are desired for the detection of CIM.
Furthermore, the current-induced moments are usually larger than the field-induced ones, since the factor $\delta/$[band width] is multiplied for the latter in $\beta$.

\begin{figure}[t!]
\begin{center}
\includegraphics[width=0.4\textwidth]{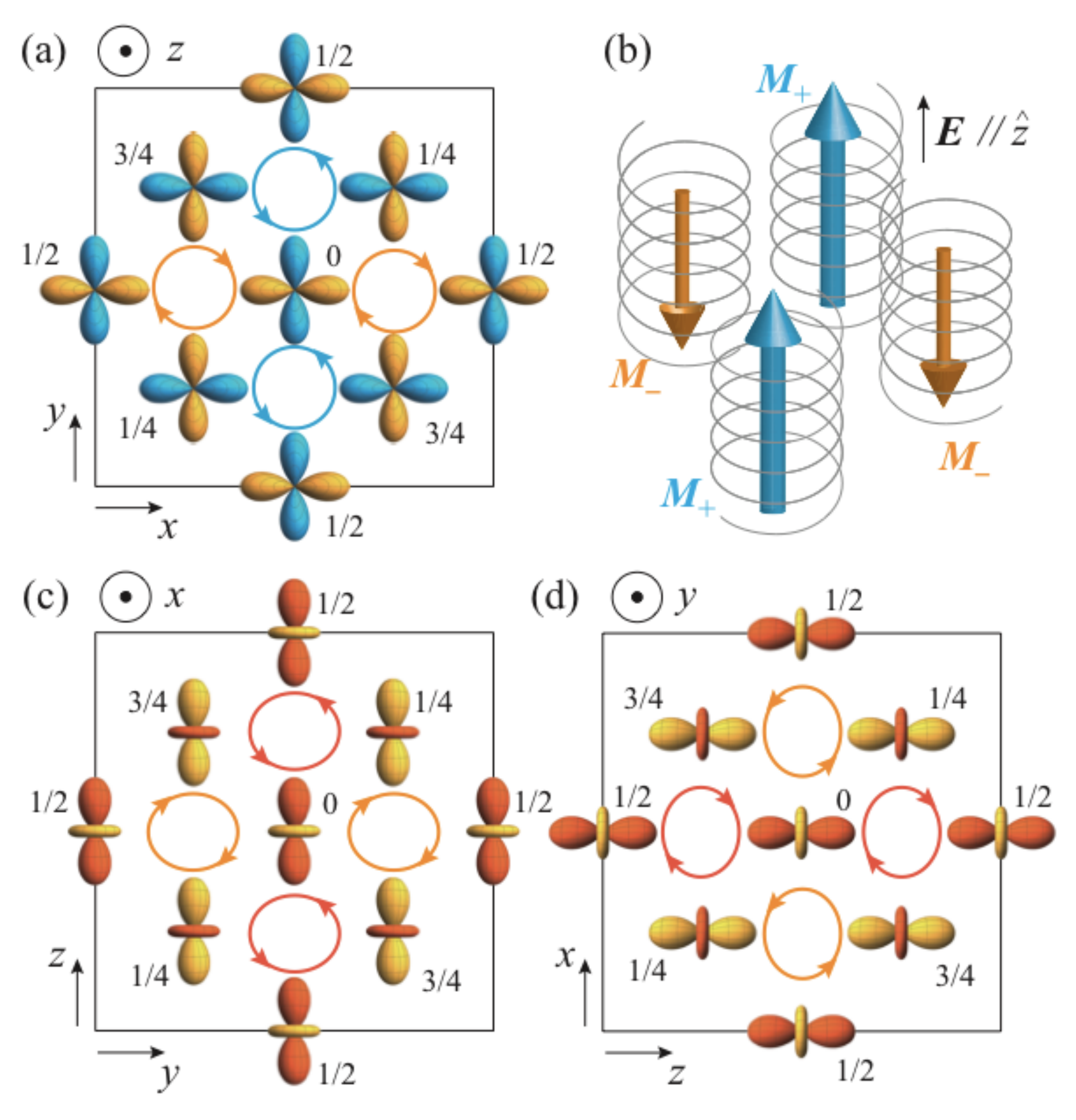}
\end{center}
\vspace{-4mm}
\caption{(color online). (a) A schematic picture for ${\mathcal O}_{22}^{\rm AF}$ order viewed from the positive $z$ direction. The color represents the sign of the charge; blue: positive charge and yellow: negative charge. The numbers indicate the coordinate along the $z$ direction and the circles with arrows represent right- and left-handed solenoids. See the main text. (b) Classical solenoid picture of the induced orbital magnetizations $\bm{M}_{+}$ and $\bm{M}_{-}$ with $|\bm{M}_{+}+\bm{M}_{-}|>0$ for (a). (c) ${\mathcal O}_{20}^{\rm AF}$ order viewed (c) from the positive $x$ direction and (d) from the positive $y$ direction. The numbers indicate the coordinate along the $x$ direction for (c) and $y$ for (d). The color represent the sign of the charge: red for positive and yellow for negative charges.}
\label{fig:solenoids}
\end{figure}

 Now, we explain that the ME effects under the AFQ orders in the diamond structure can be intuitively understood by introducing the notion of ``solenoids''.\cite{YodaYokoyamaMurakami2018} Let us first consider ${\mathcal O}_{22}^{\rm AF}$ order and view the charge distribution of the quadrupole moments from the positive $z$ direction as depicted in Fig. \ref{fig:solenoids}(a), where the positive (negative) charge are in blue (yellow) color and the numbers represent the coordinate along the $z$-direction. In the following discussion, it is better to regard the ${\mathcal O}_{22}$ moment as a complex of two positive (blue) and two negative (yellow) charges, where the sign is determined by ${\mathcal O}_{22}\propto x^2-y^2$. 
Starting from the positive (negative) charge at the $z=0$ site and follow the same charge at the nearest-neighbor site in such a way that $z$ increases by $1/4$. Resulting trajectory forms a counter-clockwise, i.e., right-handed (clockwise: left-handed) solenoid when the positive (negative) charges are traced. As depicted in Fig. \ref{fig:solenoids}(a), these solenoids align along the $z$ direction and the sign of the charges for the neighboring solenoids are opposite. This in addition to the atomic spherical charge distributions generates charge imbalance between the two kinds of solenoids. Suppose electrons do not hop among different solenoids for $\bm{E}\parallel [001]$, the resulting current along the two solenoids are different in magnitude. This is owing to, e.g., the Coulomb interactions. As a result, a residual net orbital magnetization $\bm{M}^{\rm orb}_{z}=\bm{M}_{+}+\bm{M}_{-}$ remains as depicted in Fig. \ref{fig:solenoids}(b).

%%%%%%%%%%%%%%%%%%%%%%%%%%%%%%%%%%%%%%%%%%%%%%%%%%%%%%%%%%

%%%%%%%%%%%%%%%%%%%%%%%%%%%%%%%%%%%%%%%%%%%%%%%%%%%%%%%%%%

This simple intuitive view can also explain the anisotropy in the ME effects. Consider next ${\mathcal O}_{20}^{\rm AF}$ order. In Figs. \ref{fig:solenoids}(c) and \ref{fig:solenoids}(d), ${\mathcal O}_{20}^{\rm AF}$ order  is drawn from (c) positive $x$ and (d) $y$ directions. This time, right-handed solenoids are positively charged (red) along the $x$ direction, while they are negatively charged (yellow) along the $y$ direction. This readily indicates that $\alpha_{\parallel}^{\rm orb}$ is opposite in the $x$ and $y$ directions in complete agreement with the symmetry argument in Eq. (\ref{eq:lambda2}).

In the above classical picture, it seems crucial to introduce two separated solenoids leading to a finite magnetization. In the microscopic lattice model in the diamond structure, the path for a right-handed solenoid is shared by the neighboring left-handed solenoid. Readers might wonder the validity of the solenoid interpretation. 
The microscopic origin of the ME effects in our model arises from the redistribution of $\bm{M}^{J}_{nm,\bm{k}}$ and $\bm{m}_{n\bm{k}}$ under the AFQ orders. This is demonstrated in Figs. \ref{fig:angle-dep}(b) and (c), for which the orbital dependent hoppings together with the spin-orbit coupling integrated in the $\Gamma_8$ model are important. In such microscopic mechanism, there are processes corresponding to the current flow along charge imbalanced solenoids.
Nevertheless, we note that the ME effects are related to the broken symmetry by the AFQ orders and the solenoid picture is very simple and easy to understand the essential aspect of the CIM.

In reality, there are other aspects for realizing the imbalance. Even without orbital degrees of freedom, electrons must feel the charge imbalance as in the above classical view. Furthermore, the lattice is modulated accordingly under the AFQ orders. For example, when an atom placed at the center of the nearest-neighbor bond, 16c site, exists,  this position is unstable. The atom moves on the plane perpendicular to the bond direction\cite{HattoriTsunetsugu2014}. For one solenoid, the 16c sites gather, while for the other, they move away. This readily generates charge imbalance of the conduction electrons in the two kinds of solenoids.

To summarize, we have investigated the current-induced magnetization in a diamond structure under antiferro quadrupole orders. We have found that two candidate order parameters ($\mathcal{O}_{20}^{\rm AF}, \mathcal{O}_{22}^{\rm AF}$) are distinguished by the anisotropies of the current-induced magnetization. Our results indicate that magnetoelectric effects are useful also for identification of unknown multipole orders.

\section*{Acknowledgment}
The authors thank S. Hayami, H. Kusunose, T. Taniguchi, T. D. Matsuda, and Y. Aoki for fruitful discussions.
This work was supported by a Grant-in-Aid for Scientific Research (Grant Nos. 16H01079, 16H04017, 18K03522) from the Japan Society for the Promotion of Science.

%\bibliography{Ref}

\end{document}